\documentclass[runningheads,a4paper]{llncs}

\usepackage{booktabs}
\usepackage{makecell}
\usepackage{multirow}
\usepackage{pbox}
\usepackage{subfigure}
\usepackage{times}
\usepackage{tabu}
\usepackage{tabularx}
\usepackage{epigraph}
\usepackage{tabularx}
\usepackage{verbatim}
\usepackage{url}
\usepackage{amsmath}

\usepackage{amssymb,bm}
\usepackage{sansmath}
\usepackage{mathtools}
\usepackage{float}
\usepackage{footmisc}
\usepackage{lipsum}
\usepackage{booktabs}
\usepackage{breqn}
\usepackage{graphicx,epstopdf}
\usepackage[flushleft]{threeparttable}
\usepackage{wrapfig}
\newcommand{\norm}[1]{\ensuremath{\lVert{#1}\rVert}}

\usepackage[ruled]{algorithm2e}

\urldef{\mailsa}\path|{alfred.hofmann, ursula.barth, ingrid.haas, frank.holzwarth,|
\urldef{\mailsb}\path|anna.kramer, leonie.kunz, christine.reiss, nicole.sator,|
\urldef{\mailsc}\path|erika.siebert-cole, peter.strasser, lncs}@springer.com|

\begin{document}

\mainmatter  

\title{Integrating Reviews into Personalized Ranking for Cold Start Recommendation}

\titlerunning{Integrating Reviews into Personalized Ranking}

\author{Guang-Neng Hu$^1$ \and
Xin-Yu Dai$^2$\thanks{Corresponding author} }

\authorrunning{Guang-Neng Hu and Xin-Yu Dai}

\institute{Department of Computer Science and Engineering, \\
Hong Kong University of Science and Technology, Hong Kong, China, \\
\email{njuhgn@gmail.com} \and
National Key Laboratory for Novel Software Technology, \\
Nanjing University, Nanjing 210023, China \\
\email{daixinyu@nju.edu.cn} }
%

\maketitle

\begin{abstract}
Item recommendation task predicts a personalized ranking over a set of items for each individual user. One paradigm is the rating-based methods that concentrate on explicit feedbacks and hence face the difficulties in collecting them. Meanwhile, the ranking-based methods are presented with rated items and then rank the rated above the unrated. This paradigm takes advantage of widely available implicit feedback. It, however, usually ignores a kind of important information: item reviews. Item reviews not only justify the preferences of users, but also help alleviate the cold-start problem that fails the collaborative filtering. In this paper, we propose two novel and simple models to integrate item reviews into Bayesian personalized ranking. In each model, we make use of text features extracted from item reviews using word embeddings. On top of text features we uncover the review dimensions that explain the variation in users' feedback and these review factors represent a prior preference of users. Experiments on six real-world data sets show the benefits of leveraging item reviews on ranking prediction. We also conduct analyses to understand the proposed models.

\end{abstract}

\section{Introduction}

Users confront with the ``information overload'' dilemma and it is increasingly difficult for them to choose the preferred items over others because of the growing large item set, e.g., hundreds of millions products at Amazon.com~\cite{linden03:amazon}, tens of thousands videos at Netflix.com~\cite{netflix07}. Recommender systems (RSs) assist users in tackling this problem and help them make choices by ranking the items based on their past history behavior. Item recommendation predicts a personalized ranking over a set of items for individual user and hence leads to personalized recommendation.

The rating-based (or point-wise) methods predict ratings that a user will give to items and then rank the items according to their predicted ratings. Several methods are proposed and matrix factorization based models are most popular due to their scalability, simplicity, and flexibility~\cite{PMF,SVDPP,PureSVD}. This paradigm concentrates on explicit feedback and it faces the difficulties in collecting them. Meanwhile, the ranking-based (pair-wise) methods are presented with seen items and then rank the seen above the unseen. Bayesian personalized ranking (BPR-MF) and collaborative item selection are typical representatives~\cite{BPR,CIS}. This paradigm takes advantage of widely available implicit feedback but it usually ignores a kind of important information: item reviews.

\textbf{Related works. \quad} Item reviews justify the preferences of users and help alleviate the cold-start problem. It is a diverse and complementary data source for recommendation beyond the user-item co-rating information. The collective matrix factorization (CMF) method~\cite{CMF} can be adapted to factorize the item-word matrix as well as the user-item matrix. The collaborative topic regression (CTR)~\cite{CTR} and (hidden factors and topics (HFT)~\cite{HFT} models integrate user-item interactions with item text content to build better rating predictors. They both employ topic modeling to learn hidden topic factors which explain the variations of users' preferences. The CTRank model~\cite{CTRank} also adopts topic modeling to exploit item meta-data like article titles and abstracts using bag-of-words representation for one-class CF~\cite{OCCF}. The CDR~\cite{CDR} and CKE~\cite{CKE} use deep learning techniques or neural networks such as stacked denoising autoencoders to mine the text content. Nevertheless, integrating item reviews into the ranking-based methods presents both opportunities and challenges for traditional BPR. There are few works on leveraging item reviews to improve personalized ranking. Beyond reviews, other auxiliary sources such as social relations are also intergraded into CF models~\cite{mr3}. We focus on the item reviews.

In this paper we propose two novel and simple models to incorporate item reviews into matrix factorization based Bayesian personalized ranking. Like HFT, they integrate item reviews and unlike HFT they generate a ranked list of items for individual ranking. Like CTRank, they focus on personalized ranking and unlike CTRank they are based on matrix factorization and using word embeddings to extract features. Like BPR-MF, they rank preferred items over others and unlike BPR-MF they leverage the information from item reviews. In each of the two models, we make use of text features extracted from item reviews using word embeddings. And on top of text features we uncover the review dimensions that explain the variation in users' feedback. These review factors represent a prior preference of a user. One model treats the review factor space independent of the latent factor space; another connects implicit feedback and item reviews through the shared item space.

The contributions of this work are summarized as follows.

1. We propose two novel models to integrate item reviews into matrix factorization based Bayesian personalized ranking (Section~\ref{paper:different} and Section~\ref{paper:shared}). They generate a ranked list of items for individual user by leveraging the information from item reviews.

2. For exploiting item reviews, we build the proposed models on the top of text features extracted from them. We demonstrate a simple and effective way of extracting features from item reviews by averagely composing word embeddings (Section \ref{paper:features}).

3. We empirically evaluate the proposed models on multiple real-world datasets which contains over millions of feedback in total. The experimental results show the benefit of leveraging item reviews on personalized ranking prediction. We also conduct analyses to understand the proposed models including the training efficiency and the impact of the number of latent factors.

\section{Notation and Problem Statement}

Before proposing our models, we briefly review the personalized ranking task and then describe the problem statement. To this end, we first introduce the notations used throughout the paper.

\subsection{Notation}

Suppose there are $M$ users $\mathcal{U}=\{u_1,...,u_M\}$ and $N$ items $\mathcal{I}=\{i_1,...,i_N\}$. We reserve $u,v$ for indexing users and $i,j$ for indexing items. Let $X \in \mathbb{R}^{M \times N}$ denote the user-item binary implicit feedback matrix, where $x_{u,i}$ is the preference of user $u$ on item $i$, and we mark a zero if it is unknown. Define $N_u$ as the set of items on which user $u$ has an action: $N_u \equiv \{i | i \in \mathcal{I} \land x_{u,i} > 0 \}$. Rating-based methods~\cite{PMF,SVDPP} and ranking-based methods~\cite{BPR,VBPR} are mainly to learn the latent user factors $P = [P_1,...,P_M] \in \mathbb{R}^{F \times M}$ and latent item factors $Q = [Q_1,...,Q_N] \in \mathbb{R}^{F \times N}$ from partially observed $X$.

Item $i$ may have text information, e.g., review $d_{ui}$ commented by user $u$. We aggregate all reviews of a particular item as a `doc' $d_i = \cup_{u \in \mathcal{U}} {d_{ui}}$. Approaches like CTR and HFT~\cite{CTR,HFT} integrate item content/reviews with explicit ratings for rating prediction using topic modeling. Another approach is to learn word embeddings and then compose them into document level as the item text features; we adopt this way of extracting text features $f_i \in \mathbb{R}^D$ from $d_i$ (see Section~\ref{paper:features}).

\subsection{Problem Statement}\label{paper:problem}

Our work focuses on the item recommendation or personalized ranking task where a ranked list of items is generated for each individual user. The goal is to accurately rank the unobserved items which contain both truly negative items (e.g., the user dislikes the Netflix movies or is not interesting in buying Amazon products) and missing ones (e.g., the user wants to see a movie or buy a product in the future when she knows it).

Instead of accurately predicting unseen ratings by learning a model from training samples $(u,i,x_{u,i})$ where $x_{u,i} > 0$, personalized ranking optimizes for correctly ranking item pairs by learning a model from training samples $D_S \equiv \{ (u,i,j)| u \in \mathcal{U} \land  i \in N_u \land j \in \mathcal{I} \backslash N_u \}$. The meaning of item pairs of a user $(u,i,j)$ is that she prefers the former than the latter, i.e., the model tries to reconstruct parts of a total order $>_u$ for each user $u$. From the history feedback $X$ we can infer that the observed items $i$ are ranked higher than the unobserved ones $j$; and for both observed items $i_1,i_2$ or both unobserved items $j_1,j_2$ we can infer nothing. Random (negative) sampling is adopted since the number of such pairs is huge. See the original BPR paper~\cite{BPR} for more details.

\textbf{Problem 1. \quad}Personalized Ranking with Item Reviews.

\textbf{Input:} 1) A binary implicit feedback matrix $X$, 2) an item reviews corpus $C$, and 3) a user $u$ in the user set $\mathcal{U}$ .

\textbf{Output:} A ranked list $>_u$ over the unobserved items $\mathcal{I} \backslash N_u $.

In Problem 1, to generate the ranked list, we have item reviews to exploit besides implicit feedback.

\section{The Proposed Models}\label{paper:models}

In this section, we propose two models as a solution to Problem 1 which leverage item reviews into Bayesian personalized ranking. One model treats the review factor space independent of the latent factor space (Section \ref{paper:different}). Another model connects implicit feedback and item reviews through the shared item space (Section \ref{paper:shared}). In each of the two proposed models, we make use of text features extracted from item reviews using word embeddings (Section \ref{paper:features}). On top of text features we uncover the review dimensions that explain the variation in users' feedback and these review factors represent a prior preference of a user. Both models are based on basic matrix factorization (Section \ref{paper:basicMF}) and learned with Bayesian personalized ranking (Section \ref{paper:learning}).

\subsection{Basic Matrix Factorization}\label{paper:basicMF}

The basic matrix factorization (Basic MF) is mainly to find the latent user-specific feature matrix $[P_u]^M_1$ and item-specific feature matrix $[Q_i]^N_1$ to approximate the partially observed feedback matrix $X$ in the regularized least-squares (or ridge regression) sense by solving the following problem:
\begin{equation}
\label{eq:rating}
\min_{P,Q} \sum\nolimits_{x_{u,i} \neq 0} {(x_{u,i} - \hat x_{u,i})}^2 + \lambda (\norm {P}_F^2 + \norm {Q}_F^2),
\end{equation}
where $\lambda$ is the regularization parameter to avoid over-fitting. The predicted scores $\hat x_{u,i}$ can be modeled by various forms which embody the flexibility of matrix factorization. A basic form is $\hat x_{u,i}^{Basic} = \alpha + \beta_u + \beta_i + P_u^{\textrm T} Q_i$, where $\alpha$, $\beta_u$ and $\beta_i$ are biases~\cite{SVDPP}.

\subsection{Integrating Item Reviews into Basic MF: Different Space Case}\label{paper:different}

In this section, we propose our first model {\em TBPR-Diff} to integrate item reviews with implicit feedback. Analogical to the Basic MF which factorizes the ratings into user- and item- {\em latent} factors, we can factorize the reviews into user- and item- {\em text} factors (see the illustration in Figure~\ref{fig:predictors}---Up). The TBPR-Diff model sharpens this idea and teases apart the rating dimensions into latent factors and text factors:
\begin{dmath}\label{eq:diff}
\hat x_{u,i}^{Diff} = \alpha + \beta_u + \beta_i + P_u^{\mathsf{T}} Q_i + \theta_u^{\mathsf{T}} (H f_i) + \beta^{\prime^{\mathsf{T}}} f_i,
\end{dmath}
where the term $\theta_u^{\mathsf{T}} (H f_i)$ is newly introduced to capture the text interaction between user $u$ and item $i$. To exploit item reviews, text features $f_i \in \mathbb{R}^D$ are firstly extracted from item reviews using word embeddings. The embedding kernel $H \in \mathbb{R}^{K \times D}$ linearly transforms $f_i$ from text features space (e.g., 200) into a low-dimensional text rating space (e.g., 15) and then it interacts with text factors of users $\theta_u \in \mathbb{R}^K$. A text bias vector $\beta^{\prime}$ is also introduced to model users' overall preferences towards the item reviews. The details of text features extracted from item reviews using word embeddings are described later (see Section~\ref{paper:features}).

Since the text factors of users $\theta_u$ and of items $(H f_i)$ are \textit{independent} of latent factors $P_u$ and $Q_i$, there is no deep interactions between the information sources of observed feedback and item reviews, and hence they cannot benefit from each other. Also additional parameters increase the model complexity. Based on these observations, we propose another model to alleviate the above challenges.

\subsection{Integrating Item Reviews into Basic MF: Shared Space Case}\label{paper:shared}

\begin{figure}
\centering
\includegraphics[height=5.8cm,width=4in]{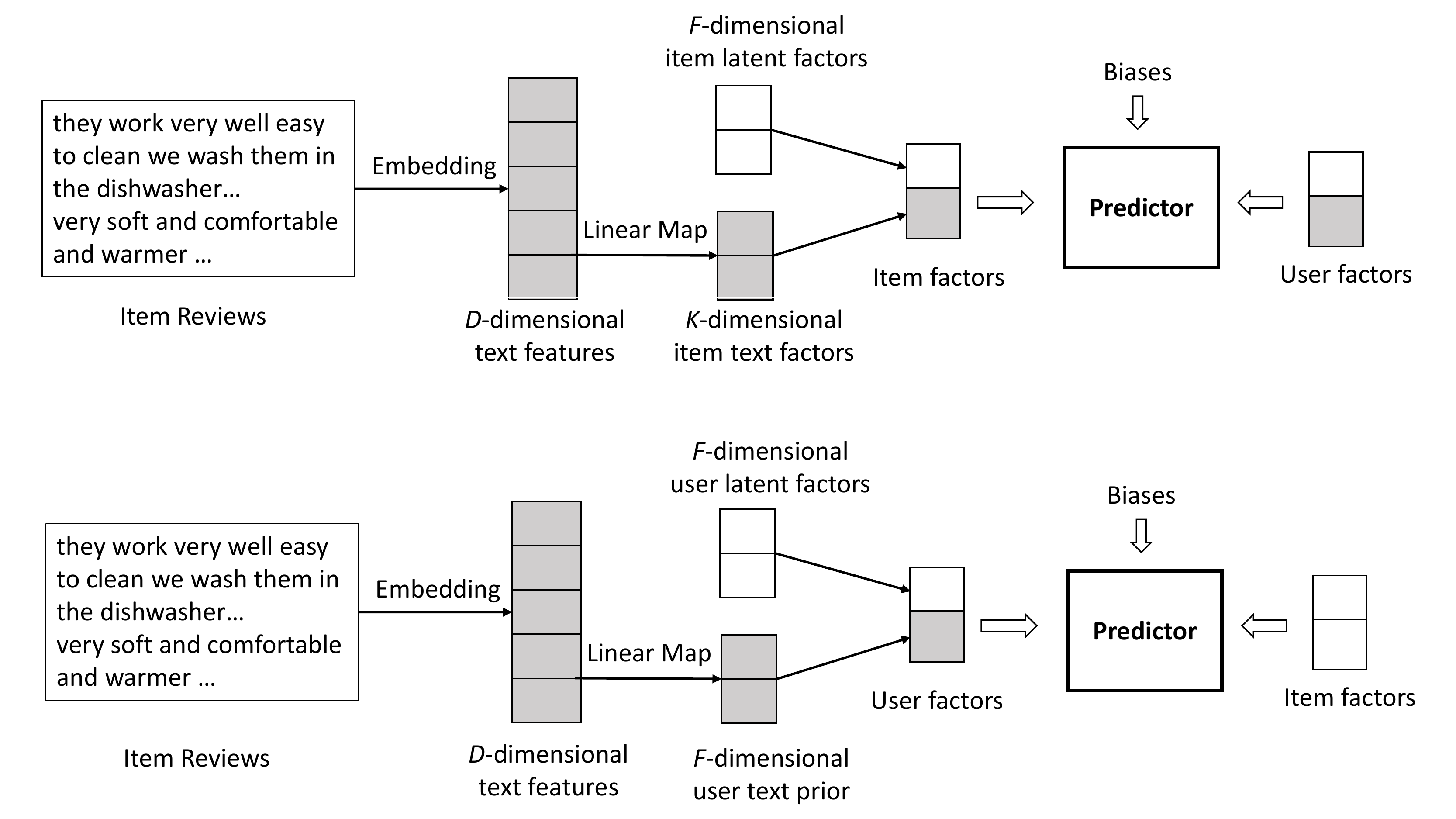}
\caption{Illustrating the Preference Predictors of Our Proposed Two Models. Up---TBPR-Diff model: The rating dimensions are to tease apart into text factors and latent factors for both user and item. Down---TBPR-Shared model: The rating dimensions are to tease apart into text factors and latent factors for only users where the text factors transformed from text features act as an a priori preference and show the `text influence of her rated items'.}
\label{fig:predictors}
\end{figure}

In this section, we propose our second model {\em TBPR-Shared} to integrate item reviews with implicit feedback more compactly. For an item $i$, its latent factors $Q_i$ learned from feedback can be considered as characteristics that it processes; meanwhile, these characteristics are probably discussed in its reviews and hence exhibit in its text factors $Hf_i$ (see the illustration in Figure~\ref{fig:predictors}---Down). For user $u$, if we let $Q_i$ and $\{Hf_k|k \in N_u\}$ be in the same space then it leads to deep interactions between text factors of user $u$ and the latent factor of item $i$. The TBPR-Shared model sharpens this idea and enables the deep interactions between text factors and latent factors as well as reduces complexity of the model:
\begin{multline}\label{eq:shared}
\hat x_{u,i}^{Shared} = Q_i^{\mathsf{T}} ( P_u + |N_u|^{-1/2} \sum\nolimits_{k \in N_u} H f_k ) + \alpha + \beta_u + \beta_i + \beta^{\prime^{\mathsf{T}}} f_i .
\end{multline}

On the right hand, the last four terms are the same with the TBPR-Diff model. Different from the TBPR-Diff model, the shared item factors $Q_i$ now have two-fold meanings: one is item latent factors that represent items' characteristics; another is to interact with item text factors that capture items' semantics from item reviews. Also different from the TBPR-Diff model, the preferences of a user now have a prior term which shows the `text influence of her rated items' captured by the text factors of corresponding items. In summary, on top of text features the TBPR-Shared model uncovers the review dimensions that explain the variation in users' feedback and these factors represent a prior preference of user.

\noindent
\textbf{Remarks \quad} \textbf{I.} The VBPR model~\cite{VBPR} proposed an analogical formulation with Eq~(\ref{eq:diff}). It exploits visual features extracted from item images and we leverage item features extracted from item reviews. The SVD++ and NSVD~\cite{SVDPP,NSVD} models proposed similar formulas with Eq~(\ref{eq:shared}). They learn an implicit feature matrix to capture implicit feedback and we learn a text correlation matrix to capture text factors; note that they didn't exploit item reviews and hence they had no the text bias term. \textbf{II.} There can be an adjustable weight on the term of text (i.e., $\theta_u^{\mathsf{T}} (H f_i)$ in Eq~(\ref{eq:diff}) and $ Q_i^{\mathsf{T}} |N_u|^{-1/2} \sum\nolimits_{k \in N_u} H f_k$ in Eq~(\ref{eq:shared})) to balance the influence from feedback and from reveiws, but here we just let feedback and reviews be equally important.

Before we delve into the learning algorithm, the preference
predictors of TBPR-Diff and of TBPR-Shared models are
shown in Figure~\ref{fig:predictors}.

\subsection{Model Learning with BPR}\label{paper:learning}

Revisit Problem 1, we need to generate a ranked list of items for individual user. Bayesian personalized ranking~\cite{BPR} is a generic pair-wise optimization framework that learns from the training item pairs using gradient descent. Denote the model parameters as $\Theta$ and let $\hat x_{uij}(\Theta)$ (for simplicity we omit model parameters and notation $x_{ui}$ is the same with $x_{u,i}$) represent an arbitrary real-valued mapping under the model parameters. Then the optimization criterion for personalized ranking BPR-OPT is:
\begin{dmath}\label{eq:bpr-opt}
\mathcal{L}(\Theta)  \equiv  \sum\nolimits_{(u,i,j) \in D_S} { \ln{\sigma(\hat x_{uij})} - \lambda \|\Theta\|^2},
\end{dmath}
where $\hat x_{uij} \equiv \hat x_{ui} - \hat x_{uj}$, and the sigmoid function is defined as $\sigma(x) =  1 / ({1 + \exp(-x)})$. The meaning behind BPR-OPT requires ranking items accurately as well as using a simple model.

Under the generic BPR-OPT framework, we derive the learning process for our proposed models TBPR-Diff and TBPR-Shared by embodying $\hat x_{ui}$ with $\hat x_{ui}^{Diff}$ and $\hat x_{ui}^{Shared}$, respectively. The BPR-OPT defined in Eq~(\ref{eq:bpr-opt}) is differentiable and hence gradient ascent methods can be used to maximize it. For stochastic gradient ascent, a triple $(u,i,j)$ is randomly sampled from training sets $D_S$ and then update the model parameters by:
\begin{dmath}\label{eq:update-sgd}
\Theta \leftarrow \Theta + \eta ( \sigma(-\hat x_{uij}) \frac{\partial \hat x_{uij}}{\partial \Theta} - \lambda \Theta ).
\end{dmath}

The same gradients for user latent factors and bias terms of both models are:
\begin{equation*}\label{eq:grad-P}
\frac{\partial}{\partial P_u} \hat x_{uij} = Q_i - Q_j, \; \frac{\partial}{\partial \beta^{\prime}} \hat x_{uij} = f_i - f_j, \; \frac{\partial}{\partial \beta_i} \hat x_{uij} = 1, \; \frac{\partial}{\partial \beta_j} \hat x_{uij} = -1.
\end{equation*}

Parameter gradients of the model TBPR-Diff are:
\begin{equation*}
\frac{\partial}{\partial Q_i} \hat x_{uij} = P_u, \; \frac{\partial}{\partial Q_j} \hat x_{uij} = - P_u, \; \frac{\partial}{\partial \theta_u} \hat x_{uij} = H (f_i - f_j), \; \frac{\partial}{\partial H} \hat x_{uij} = \theta_u (f_i - f_j)^{\mathsf{T}}.
\end{equation*}

Parameter gradients of the model TBPR-Shared are:
\begin{equation*}
\frac{\partial}{\partial Q_i} \hat x_{uij} = P_u + |N_u|^{-1/2} \sum\nolimits_{k \in N_u} H {f_k}, \; \frac{\partial}{\partial Q_j} \hat x_{uij} = - ( P_u + |N_u|^{-1/2}  \sum\nolimits_{k \in N_u} H {f_k}),
\end{equation*}

\begin{equation*}
\frac{\partial}{\partial H} \hat x_{uij} = |N_u|^{-1/2}  (Q_i - Q_j) ({\sum\nolimits_{k \in N_u} {f_k}})^{\mathsf{T}}.
\end{equation*}

\textbf{Complexity of Models and Learning. \quad} The complexity of model TBPR-Diff is $(M + N)F + (M + D)K + D$ while the complexity of model TBPR-Shared is $(M + N)F + (D + 1)K$. We can see that the latter model reduces the complexity by $\mathcal{O}(MK)$, i.e., the parameters $[\theta_u]^M_1$. For updating each training sample $(u,i,j) \in D_S$, the complexity of learning TBPR-Diff is linear in the number of dimensions ($F,K,D$) while the complexity of learning TBPR-Shared is also linear provided that the scale of rated items of users is amortizing constant, i.e., $\sum_{u \in \mathcal{U}} {|N_u|/|\mathcal{U}|} \approx const \ll |\mathcal{I}|$, which holds in real-world datasets because of sparsity (see Table~\ref{table:datasets}).

\section{Feature Representations of Item Reviews}\label{paper:features}

Recall that when generating the ranked list of items for individual user, we have item reviews to exploit besides implicit feedback. To exploit item reviews, we extract text features from them, i.e., there is a feature vector for each item. Our proposed two models are both built on the top of text features ($[f_i]_{i=1}^{N}$) and hence they are important for improving personalized ranking. In this section, we give one simple way to extract text features from reviews of item---word embedding.

The SGNS model~\cite{word2vec} is an architecture for learning continuous representations of words from large corpus; these representations, or word embeddings, can capture the syntactic and semantic relationships of words. We first run the Google word2vec code on Amazon reviews corpus (see Table~\ref{table:datasets}) using the default setting (particularly, dimensionality $D = 200$) to learn a vector $\mathbf{e}_w$ for each word $w$. And then we directly sum up all of the embeddings in an item's reviews (excluding stop words) and get a composition vector as the text feature for this item:
\begin{equation}\label{eq:embedding}
f_i \equiv \frac{1}{|d_i|}\sum\nolimits_{w \in d_i} {\mathbf{e}_w}.
\end{equation}
To get $f_i$, we can also use complex methods to compose the individual embeddings~\cite{neuraltensor} and to learn the doc representation directly~\cite{SDAE}; these complex methods are left for future investigation.

\section{Experiments}\label{paper:experiments}

We have proposed two models towards a solution to the Problem 1. The two models TBPR-Diff and TBPR-Shared integrate item reviews into Bayesian personalized ranking optimization criterion and uncover the text dimensions in users' feedback. We want to know the benefit of leveraging item reviews and so we compare them with BPR-MF~\cite{BPR} which ignores the information of item reviews. In addition we report the results for the most popular (POP) baseline that predicts item pairs by their corresponding `popularity' and this method doesn't show personalized ranking. Furthermore, we analyse the impact of the number of latent factors on our proposed models.

\begin{table}
\centering
\caption{Statistics of Datasets}
\label{table:datasets}
\resizebox{\columnwidth}{!}{%
\begin{tabular}{c| rrrrrr| c}
\Xhline{2\arrayrulewidth}
\hline
Datasets &\#Users &\#Items & \#Feedback & \#Words & \#Cold Users & \#Cold Items & Density (\%)\\
\hline \hline
Girls &778    &3,963   &5,474    &302M &572    &3,946  & 0.177  \\
Boys  &981    &4,114   &6,388    &302M &787    &4,080  & 0.158  \\
Baby  &1,238  &4,592   &8,401    &302M &959    &4,482  & 0.147   \\
Men   &21,793 &55,647  &157,329  &302M &15,821 &52,031 & 0.013  \\
Women &62,928 &157,656 &504,847  &302M &41,409 &143,444& 0.005 \\
Phones&58,741 &77,979  &420,847  &210M &43,429 &67,706 & 0.009  \\
\Xhline{2\arrayrulewidth}
\end{tabular}%
}
\end{table}

\subsection{Datasets}\label{paper:data}

We evaluate our models on six Amazon datasets\quad\;~\url{http://jmcauley.ucsd.edu/data/amazon/}. They consist of five from clothing and shoes category, and one from cell phones and accessories. We use the review history as implicit feedback and aggregate all users' reviews to an item as a doc for this item. We draw the samples from original datasets such that every user has rated at least five items (i.e., $\forall u \in \mathcal{U}: |N_u| \geq 5$) and the statistics of final evaluation datasets are show in Table~\ref{table:datasets}. From the table we can see that: 1) the observed feedback is very sparse, typically less than 0.01\%; 2) the average feedback events for users are typical about ten, i.e., $\sum_{u \in \mathcal{U}} {|N_u|/|\mathcal{U}|} \approx 10 \ll |\mathcal{I}|$ holds; 3) more than half of the users and of the items are cold and have feedback less than seven. Note that the cold-users/-items are those that have less than seven feedback events, and the feedback Density = $\#Feedback / (\#Users * \#Items)$.

We split each of the whole datasets into three parts: training, validation, and test. In detail, for each user $u \in \mathcal{U}$, we randomly sample two items from her history feedback for test set $Test_u$, two for validation set $Valid_u$, and the rest for training set $Train_u$; and hence $N_u = Train_u \cup Valid_u \cup Test_u$. This is the reason that we discard users who rated items less than five to ensure that there is at least one training sample for her.

\subsection{Evaluation Protocol}

For item recommendation or personalized ranking, we need to generate a ranked list over the unobserved items. Therefore for the hold-out test item $i \in Test_u$ of individual user $u$, the evaluation calculates how accurately the model rank $i$ over other unobserved items $j \in \mathcal{I} \backslash N_u$. The widely used measure Area Under the ROC Curve (AUC) sharpens the ranking correctness intuition:

\begin{equation}\label{eq:auc-bpr}
AUC = \frac {1} {|\mathcal{U}|} \sum\limits_{u \in \mathcal{U} } {\frac {1} {|E(u)|}} \sum\limits_{(i,j) \in E(u)}{\delta{(\hat x_{u,i} > \hat x_{u,j})}},
\end{equation}
where $E(u) = \{(i,j) | i \in Test_u  \land j \in \mathcal{I} \land j \notin N_u \}$ and the $\delta(\cdot)$ is an indicator function. A higher AUC score indicates a better recommendation performance.

The validation set $\mathcal{V} = \cup_{u \in \mathcal{U}} Valid_u$ is used to tune hyperparameters and we report the corresponding results on the test set $\mathcal{T} = \cup_{u \in \mathcal{U}} Test_u$.

\begin{table}
\centering
\caption{AUC Performance Results (\#factors = 15, best result is {\textbf{boldfaced}}).}
\label{table:results}
\resizebox{\columnwidth}{!}{%
\begin{tabular}{c| ccccc|cc }
\Xhline{2\arrayrulewidth}
\hline
Datasets&Setting& POP  & BPR-MF & TBPR-Diff&TBPR-Shared& Improv1 (\%) & Improv2 (\%)\\
\hline \hline
Girls  & All  & 0.1699 & 0.5658 & 0.5919 & \textbf{0.5939} & 4.966 & 7.09\\
\hline
Boys   & All  & 0.2499 & 0.5493 & 0.5808 & \textbf{0.5852} & 6.535 & 11.99\\
\hline
Baby   & All  & 0.3451 & 0.5663 & 0.5932 & \textbf{0.6021} & 6.321 & 16.18\\
\hline
       & All  & 0.5486 & 0.6536 & 0.6639 & \textbf{0.6731} & 2.983 & 18.57\\
Men    & Cold & 0.4725 & 0.5983 & 0.6114 & \textbf{0.6225} & 4.044 & 19.23\\
\hline
       & All  & 0.5894 & 0.6735 & 0.6797 & \textbf{0.6842} & 1.588 & 12.72\\
Women  & Cold & 0.4904 & 0.6026 & 0.6110 & \textbf{0.6152} & 2.090 & 11.22\\
\hline
       & All  & 0.7310 & 0.7779 & 0.7799 & \textbf{0.7809} & 0.386 & 6.39\\
Phones & Cold & 0.5539 & 0.6415 & 0.6464 & \textbf{0.6467} & 0.811 & 5.94 \\
\Xhline{2\arrayrulewidth}
\end{tabular}%
}
\end{table}

\subsection{Comparing Methods}
We compare our proposed models TBPR-Diff (see Eq~(\ref{eq:diff})) and TBPR-Shared (see Eq~(\ref{eq:shared})) with the Most Popular \textbf{(POP)} and \textbf{BPR-MF}~\cite{BPR} baselines. The difference of models lies in their preference predictors.

\noindent
{\textbf{Reproducibility}. \quad} We use the released code in~\cite{VBPR} to implement the comparing methods and our proposed models. The hyperparameters are tuned on the validation set. Referring to the default setting, for the BRP-MF model, the norm-penalty $\lambda = 11$, and learning rate $\eta = 0.005$. As with our proposed models TBPR-Diff and TBPR-Shared, the norm-penalty $\lambda_{latent} = 11$ for latent factors and $\lambda_{text} = 5$ for text factors, and learning rate $\eta = 0.001$. For simplicity, the number of latent factors equals to the number of text factors; the default values for them are both fifteen (i.e., $F = K = 15$). The impact of the number of factors is analysed in Section~\ref{paper:analysis}. Since the raw datasets, comparing code, and parameter setting are given publicly, we confidently believe our experiments are easily reproduced.

\subsection{Performance Results}

\begin{figure}
\centering
\subfigure{ \includegraphics[height=3.2cm,width=2.1in]{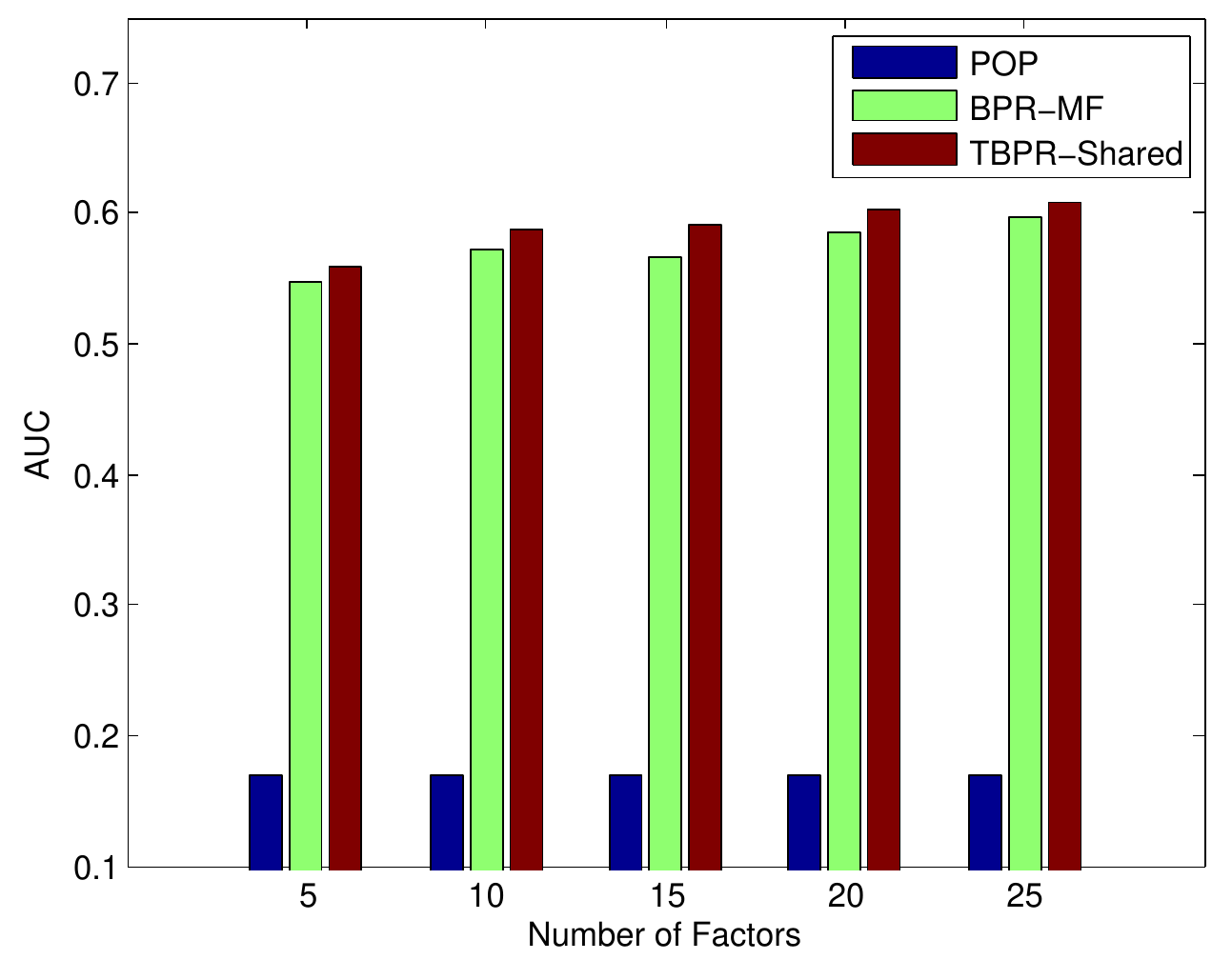}}
\subfigure{ \includegraphics[height=3.2cm,width=2.1in]{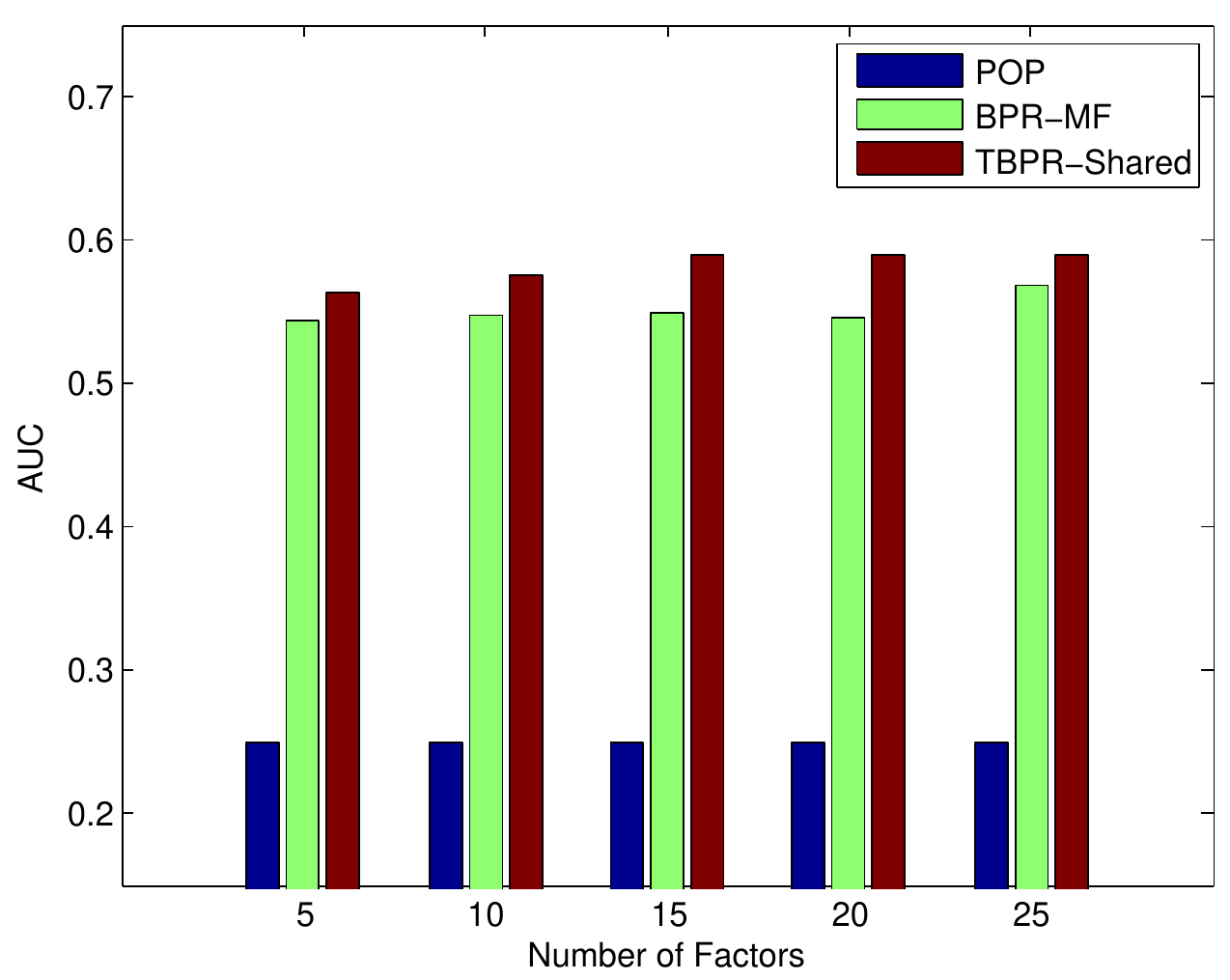}}
\subfigure{ \includegraphics[height=3.2cm,width=2.1in]{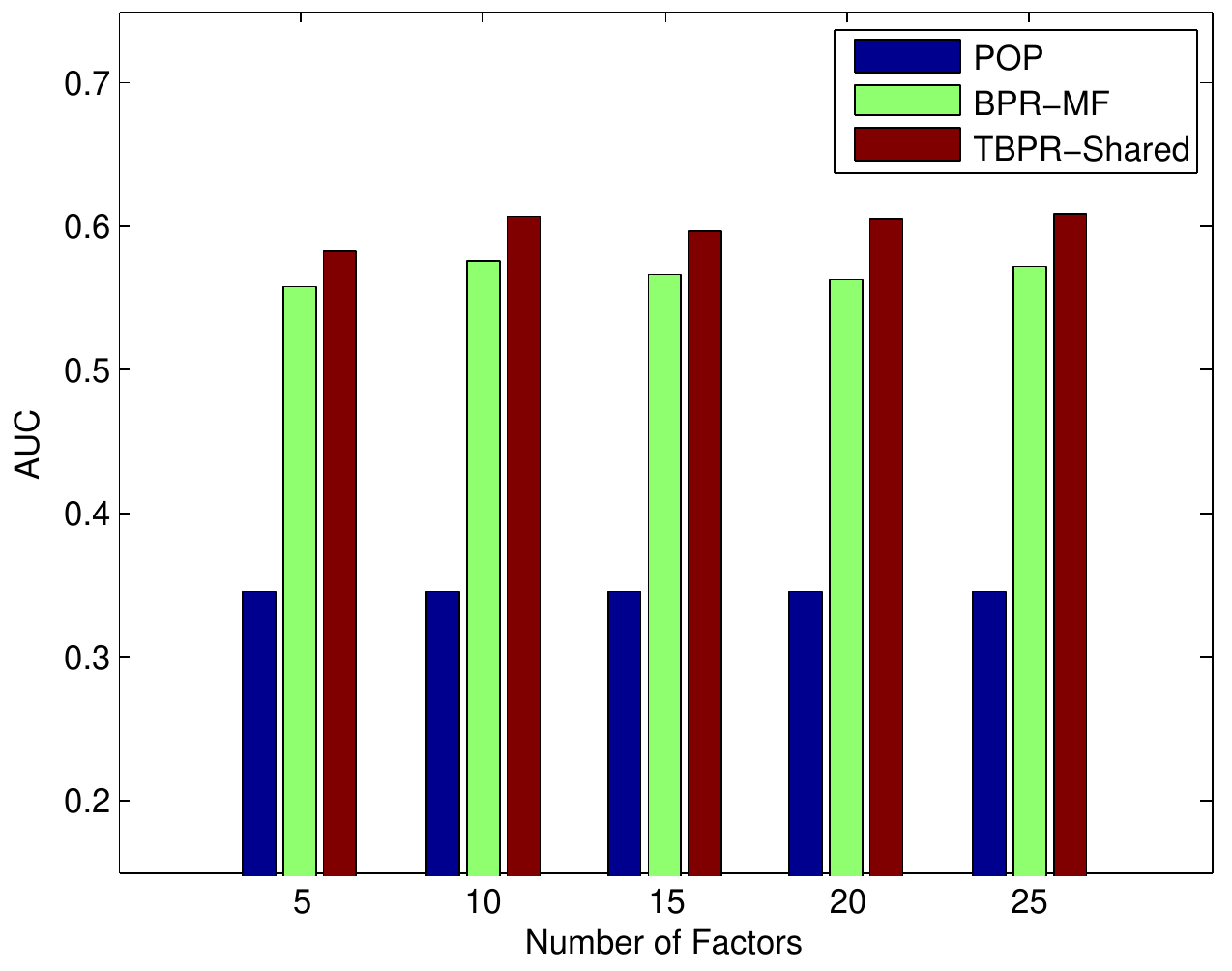} }
\subfigure{ \includegraphics[height=3.2cm,width=2.1in]{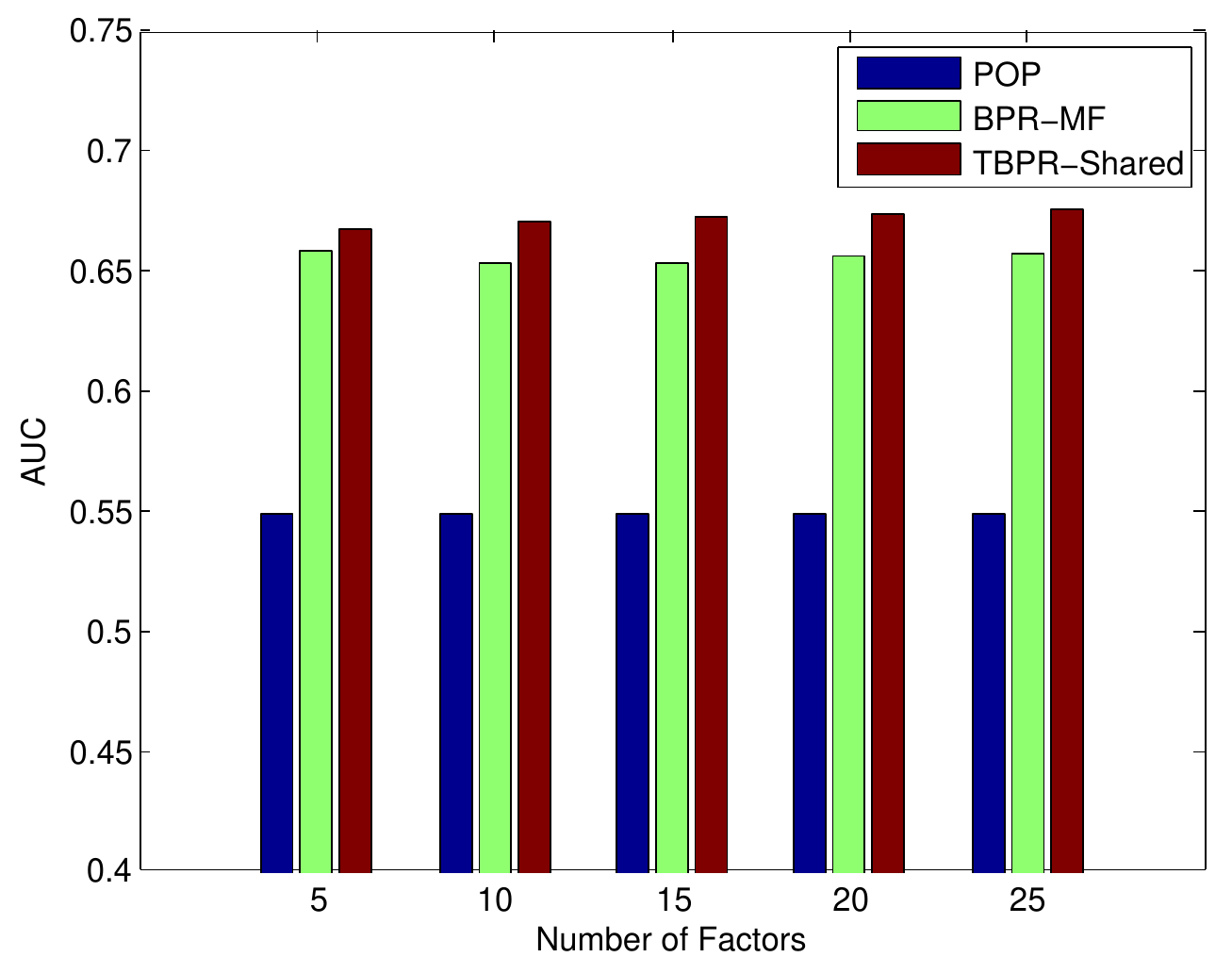}}
\caption{Performance Results of AUC on the Test Set with Varying the Number of Latent Factors. From top to bottom and left to right, the datasets are Girls, Boys, Baby, and Men (due to limited space we omit the results on Women and Phones). For clarity, we omit the results of TBPR-Diff where they are slightly better than BPR-MF and slightly worse than TBPR-Shared.}
\label{fig:factors}
\end{figure}

The AUC performance results on eight Amazon.com datasets are shown in Table~\ref{table:results} where the last but one column is $(AUC_{\mathrm{TBPR-Shared}} - AUC_{\mathrm{BPR-MF}}) / AUC_{\mathrm{BPR-MF}} \times 100\%$, and the last column is $ (AUC_{\mathrm{TBPR-Shared}} - AUC_{\mathrm{BPR-MF}}) / (AUC_{\mathrm{BPR-MF}} - AUC_{\mathrm{POP}}) \times 100\%$. For each dataset there are three evaluation settings: The {\em All Items or All} setting evaluates the models on the full test set $\mathcal{T}$; the {\em Cold Start or Cold} setting evaluates the models on a subset $\mathcal{T}_{cold} \subseteq \mathcal{T}$ such that the number of training samples for each item within $\mathcal{T}_{cold}$ is no greater than three (i.e., $|Train_u| \leq 3$ or $|N_u| \leq 7$); the {\em Warm} setting evaluates the models on the difference set of All and Cold. Revisit the Table~\ref{table:datasets} we can see that: 1) almost all of the items are cold-item for datasets Girls, Boys, and Baby; and hence the results of Cold setting are almost the same with All and the results of Warm setting is not available to get a statistical reliable results; and 2) for other three datasets, the percent of cold-items is also more than 86\% which requires the model to address the inherent cold start nature of the recommendation problem.

There are several observations from the evaluation results.

1. \textit{Under the All setting}, TBPR-Shared is the top performer, TBPR-Diff is the second, with BPR-MF coming in third and POP the weakest. These results firstly show that leveraging item reviews besides the feedback can improve the personalized ranking; and also show that the personalization methods are distinctly better than the user-independent POP method. For example, TBPR-Shared averagely obtains relative 4.83\% performance improvement compared with BPR-MF on the first three smaller datasets in terms of AUC metric, and 2.74\% in total six datasets. This two figures show, to some extent, that transferring the knowledge from auxiliary data source (here item reviews) helps most when the target data source (here rating feedback) is not so rich.

2. \textit{Under the Cold setting}, TBPR-Shared is the top performer, TBPR-Diff is the second, with BPR-MF coming in third and POP is also the weakest. These results firstly show that leveraging item reviews besides the feedback can improve the personalized ranking even in the cold start setting; and also show that the personalization methods are distinctly better than the user-independent POP method since the cold items are not popular. In detail, TBPR-Shared averagely obtains relative 2.31\% performance improvement compared with BPR-MF in terms of AUC metric. Furthermore, TBPR-Shared compared with BPR-MF, the relative improvement in the \textit{cold start setting} is about 1.6 times than that in the \textit{All setting} which implies that integrating item reviews more benefits when observed feedback is sparser. As with the results on the Phones dataset, revisiting Table~\ref{table:datasets} we can see that the ratio of cold items over all item is 86.8\% which is far less than those on other two datasets ($\sim 92.2\%$). And in this case adding auxiliary information doesn't help much.

We also evaluate on \textit{the Warm setting} (not shown in Table~\ref{table:results}), and all of the personalized, complex methods are worse than the user-independent, simple method POP. Warm items are more likely to be popular and show less personalized characteristics. It reminds us the commonplace that recommendation plays an important role in long-tailed items.

\begin{figure}
\centering
\subfigure{ \includegraphics[height=3.2cm,width=1.5in]{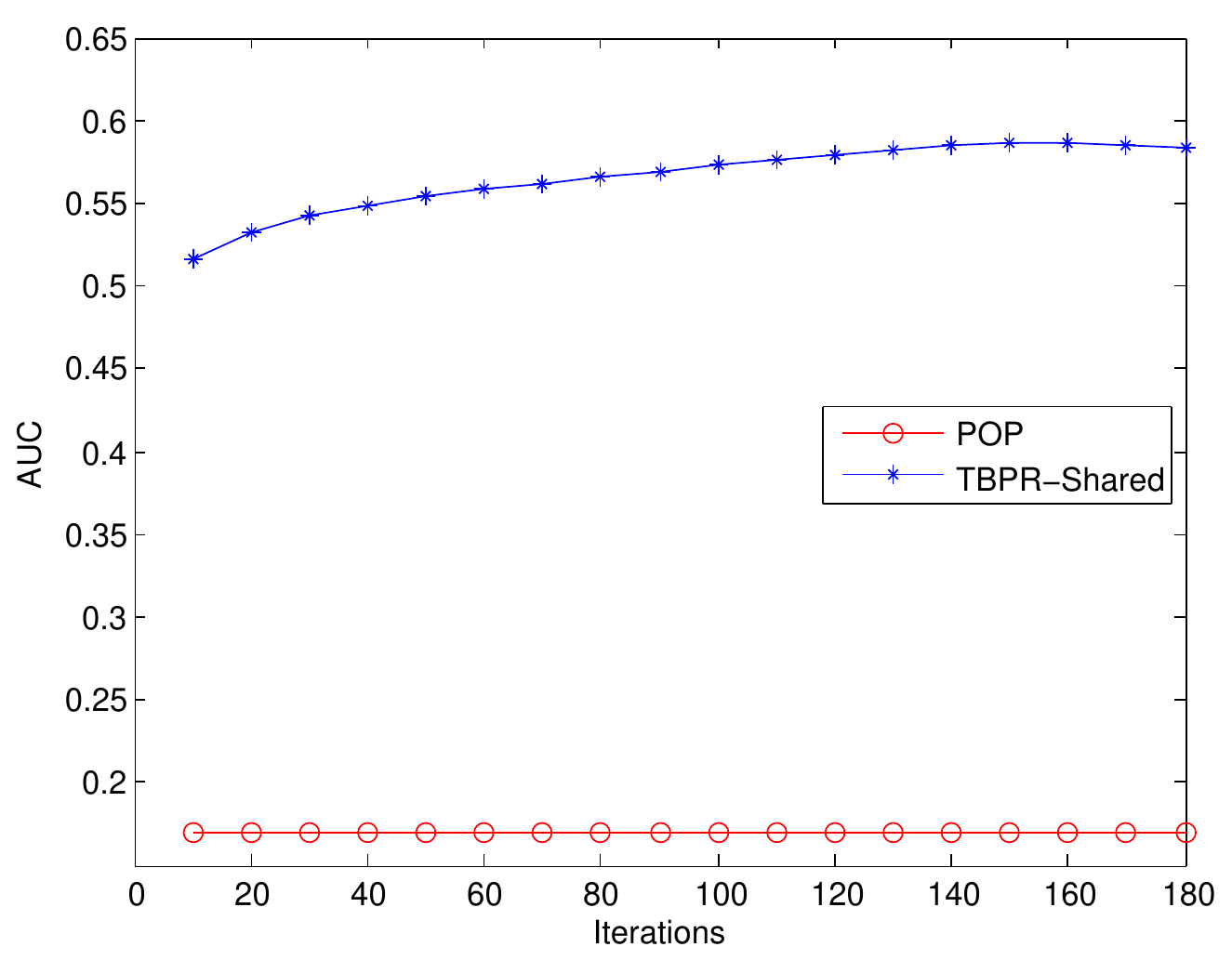} }
\subfigure{ \includegraphics[height=3.2cm,width=1.5in]{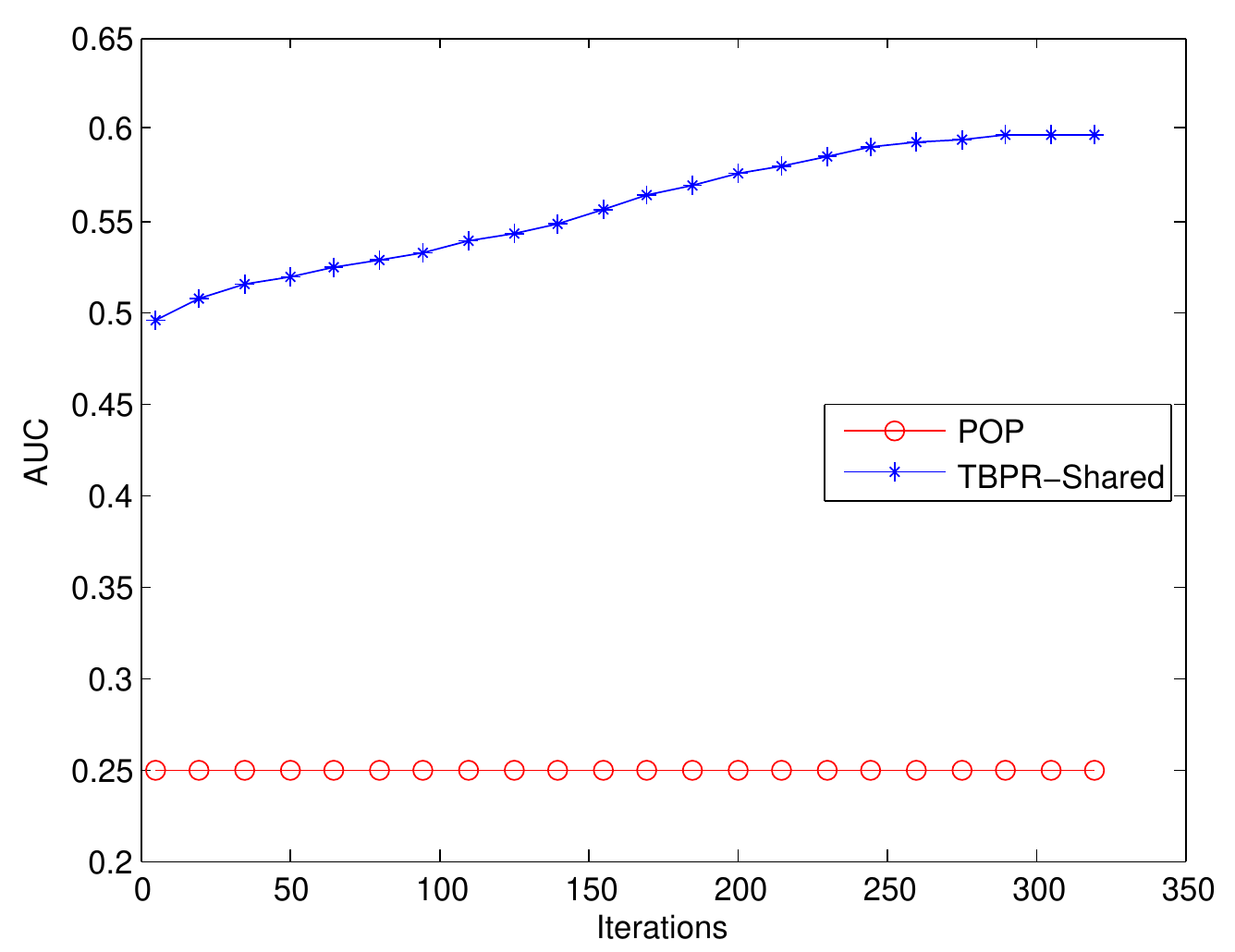} }
\subfigure{ \includegraphics[height=3.2cm,width=1.5in]{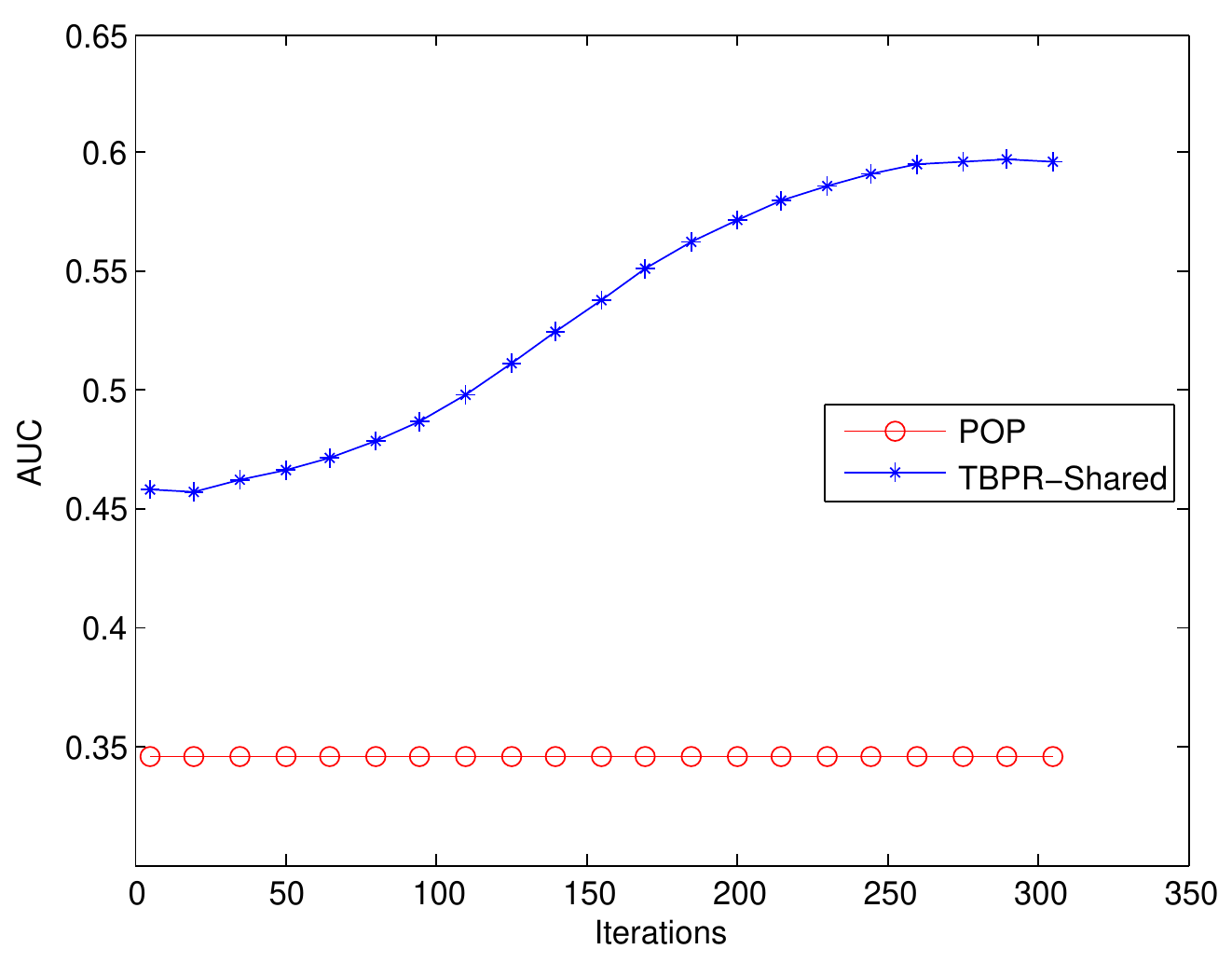} }
\subfigure{ \includegraphics[height=3.2cm,width=1.5in]{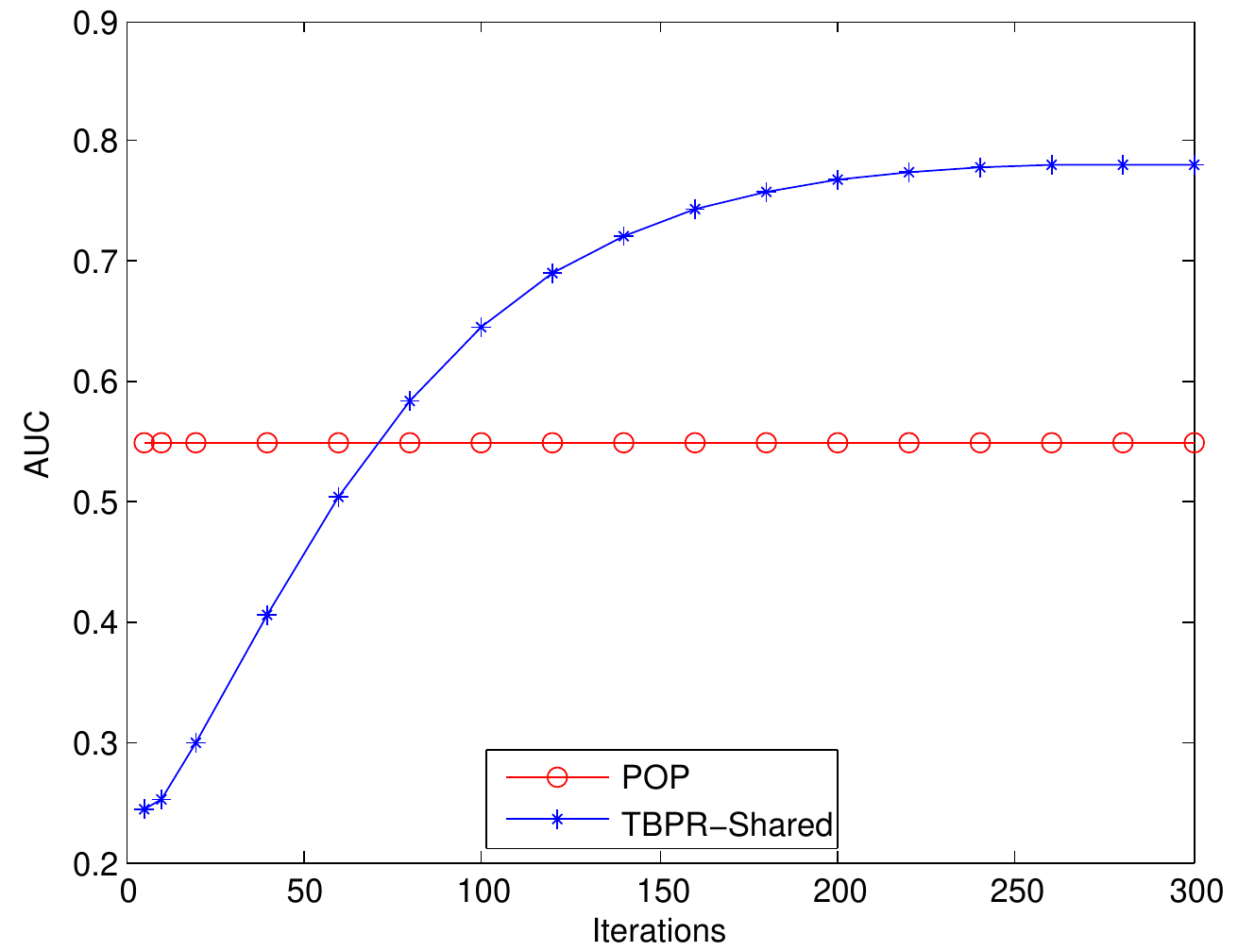} }
\subfigure{ \includegraphics[height=3.2cm,width=1.5in]{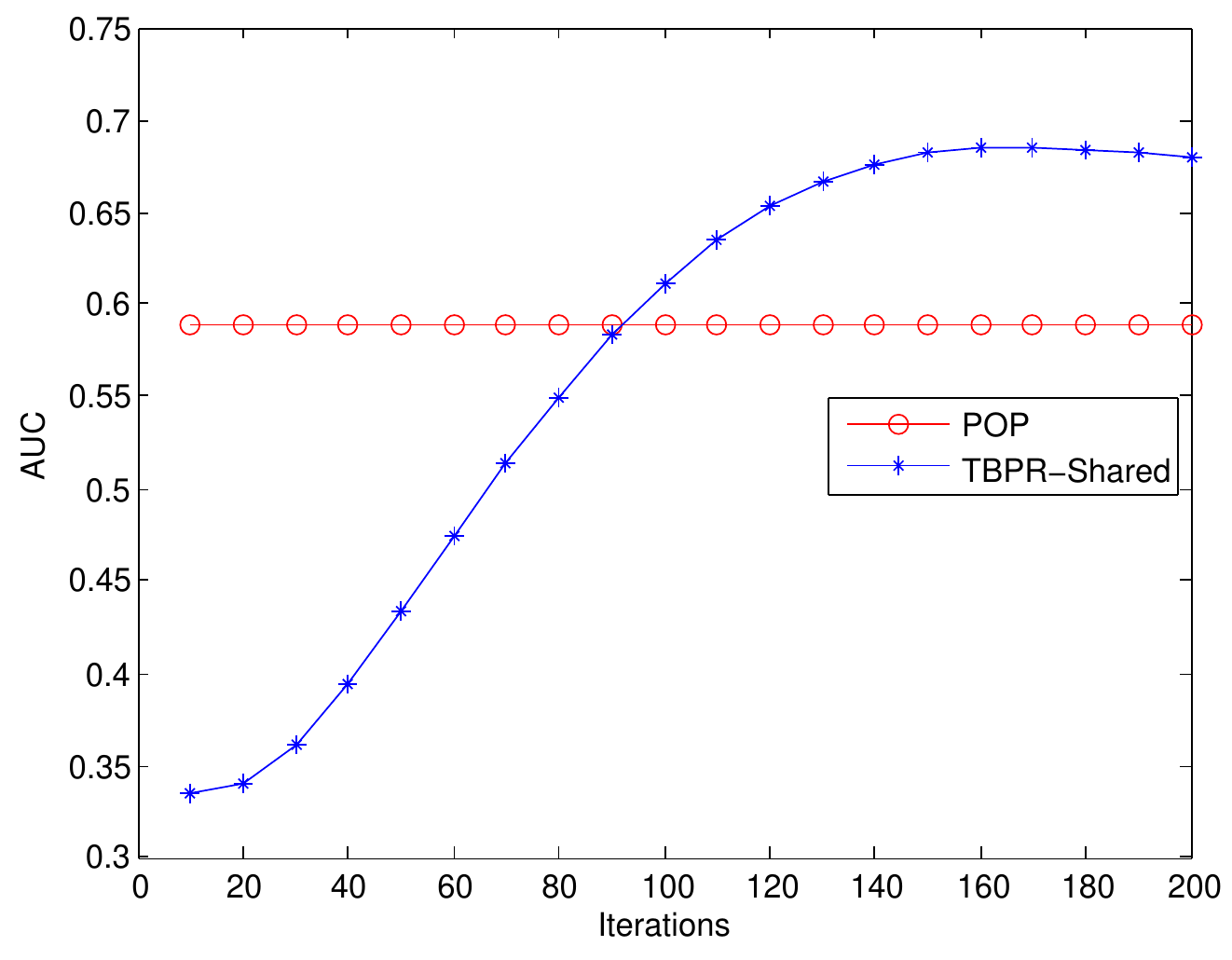} }
\subfigure{ \includegraphics[height=3.2cm,width=1.5in]{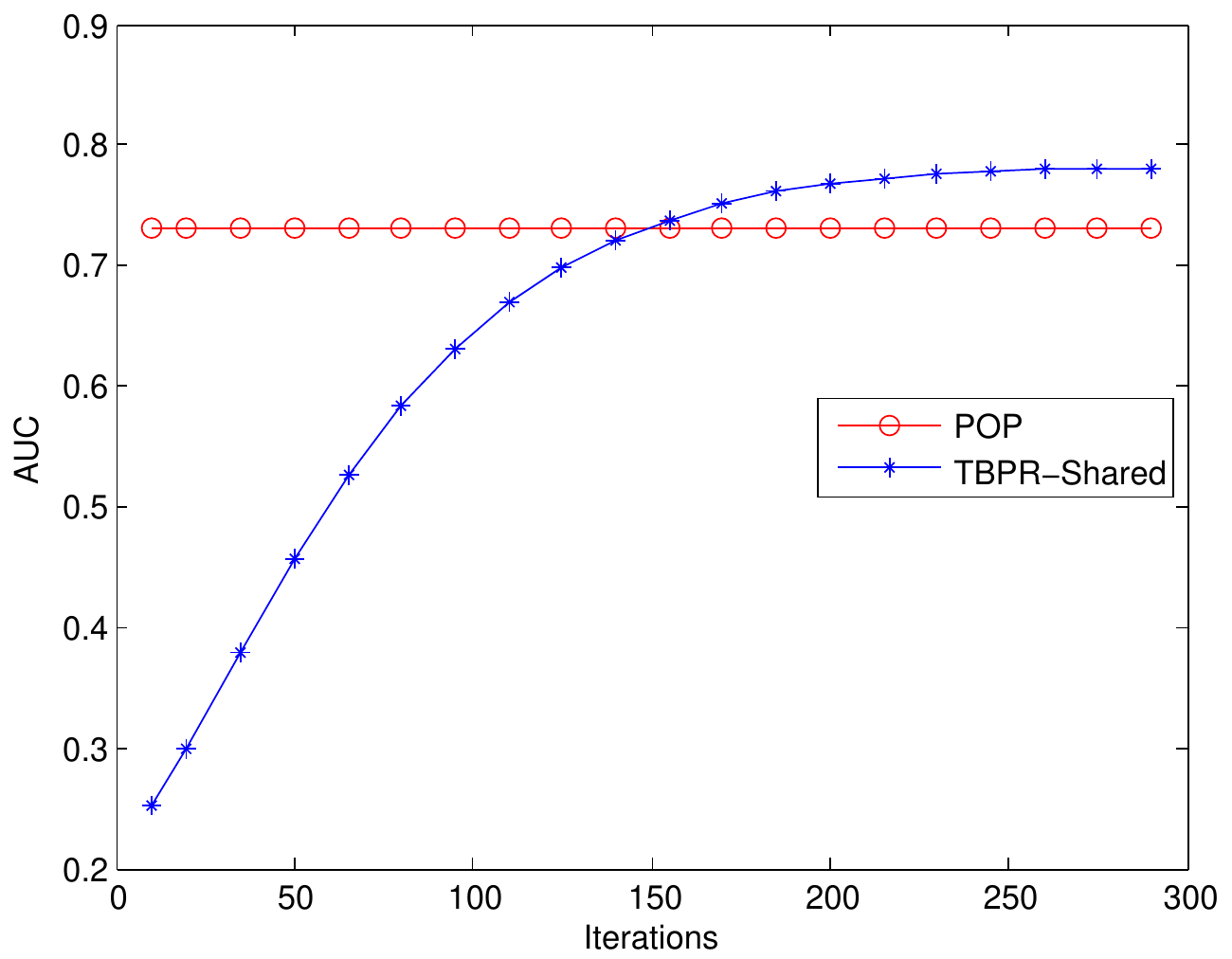} }
\caption{Performance Results of AUC on the Validation Set with Training Iterations (\#factors = 15). From top to bottom and left to right, the datasets are Girls, Boys, Baby, Men, Women, and Phones. For clarity, we omit the results of TBPR-Diff where they are slightly worse than TBPR-Shared. As a reference, BPR-MF model usually converges in 50 iterations. Due to limited space, we only give the validation results for \#factors = 15 and omit {5, 10, 20, 25}. }
\label{fig:iterations}
\end{figure}

\subsection{Analysis of the Proposed Models}\label{paper:analysis}

After demonstrating the benefits of leveraging item reviews, we analyse the proposed models from two points; one is the impact of number of latent factors, and one is the training efficiency and convergence analysis. More depth investigation like the impact of embedding dimensionality and of corpus source to train the embeddings, is left to future work.

\noindent
\textbf{Impact of the Number of Latent Factors. \quad} The two proposed models TBPR-Shared and TBPR-Diff have two important hyperparameters; one is the number of latent factors $F$ and one is the number of text factors $K$. For simplicity, we let the two values equal. We vary the number of latent factors $\#factors = \{5,10,15,20,25\}$ to observe the performance results of different methods. The test AUC scores are shown in Figure~\ref{fig:factors}. On the Girls and Boys datasets, both of the personalized models are to perform better as the number of factors increases; on the other datasets, the performance improves as the number of factors increases to around fifteen; then it doesn't go up and may even downgrade. We set the default value as 15.

Also the plots visually show the benefits of integrating item reviews (TBPR-Shared vs. BPR-MF) and of generating a personalized ranking item list for individual user (TBPR-Shared and BPR-MF vs. POP).

\noindent
\textbf{Training Efficiency and Convergence Analysis. \quad} The complexity of learning is approximately linear in the number of parameters of our proposed models. Figure~\ref{fig:iterations} shows the AUC scores of the TBPR-Shared model on validation sets with increasing training iterations. In summary, our models take 3-4 times more iterations to converge than BPR-MF. On three smaller datasets (Girls, Boys, and Baby), the first five iterations are enough to get a better score than POP; and on the other larger datasets (Men, Women, and Phones), it takes longer.

As a reference, the BPR-MF model usually converges in 50 iterations. As another reference, all of our experiments are completed in about one week using one server that has 65GiB memory and 12 cores with frequency 3599MHz.


\section{Conclusion and Future Work}

Item reviews justify the rating behavior of users and hence they are useful for improving recommender performance. Based on matrix factorization techniques we proposed two models to integrate item reviews into Bayesian personalized ranking. In each of the two models, we make use of text features extracted from item reviews using word embeddings. On top of text features we uncover the review dimensions that explain the variation in users' feedback. These review factors represent a prior preference of a user and show the `text influence of her rated items'. Empirical results on multiple real-world datasets demonstrated that our proposed models lead to improved ranking prediction performance under the All setting and the cold start setting in terms of AUC. And the shared space model is slightly better than the different space one which shows the benefits of considering the interactions between latent factors and text factors and the benefits of reducing the model complexity. Furthermore, we analyzed the impact of the dimensionality of latent factors and the efficiency of model learning.

Focusing on leveraging item reviews for improving personalized ranking, there are several directions we want to explore. First, we integrate the item reviews with implicit feedback by adding text dimensions into the rating predictors; certainly, this is not the only way to exploit item features. Second, more evaluation metrics should be explored besides AUC (e.g., hit rate). Third, the construction strategy of positive/negative samples is also worth further investigating because it deeply affects the modeling design, the learning results, and the evaluation performance. Last but not the least, since we investigate the benefits of leveraging item reviews, we only compare our models with BPR-MF (and POP); and to know the effectiveness, comparing with more baselines is needed.

\noindent
{\bf Acknowledgments } The work is supported by HKPFS (PF15-16701), NSFC (61472183), and 863 Program (2015AA015406).


\end{document}